\documentclass{ws-procs9x6} 
 
\usepackage{graphicx} 

\renewcommand{\theenumi}{\arabic{enumi}}

\begin{document} 
 
\title{Aspects of Dynamical Chiral Symmetry Breaking} 
 
\author{C.~D.\ Roberts} 
 
\address{Physics Division, Argonne National Laboratory\\ Argonne, 
Illinois, 60439-4843, USA} 
 
 
\maketitle 
 
\abstracts{Dynamical chiral symmetry breaking is a nonperturbative phenomenon
that may be studied using QCD's gap equation.  Model-independent results can
be obtained with a nonperturbative and symmetry preserving truncation.  The
gap equation yields the massive dressed-quark propagator, which has a
spectral representation when considered as a function of the current-quark
mass.  This enables an explication of the connection between the infrared
limit of the QCD Dirac operator's spectrum and the quark condensate appearing
in the operator product expansion.}
 
\section{Introduction} 
Dynamical chiral symmetry breaking (DCSB) is fundamental to the strong
interaction spectrum.  It is the generation through self-interactions of a
momentum-dependent quark mass, $M(p^2)$, in the chiral limit, that is large
in the infrared but power-suppressed in the ultraviolet:
\begin{equation} 
\label{Mchiral} 
M(p^2) \stackrel{{\rm large}-p^2}{=}\, 
\frac{2\pi^2\gamma_m}{3}\,\frac{\left(-\,\langle \bar q q \rangle^0\right)} 
{p^2 \left(\frac{1}{2}\ln\left[\frac{p^2}{\Lambda_{\rm QCD}^2}\right] 
\right)^{1-\gamma_m}}\,. 
\end{equation} 
The phenomenon is impossible in weakly interacting theories.  In Eq.\
(\ref{Mchiral}), $\gamma_m=12/(33-2N_f)$ is the mass anomalous dimension,
with $N_f$ the number of light-quark flavours, and $\langle \bar q q
\rangle^0$ is the renormalisation-group-invariant vacuum quark
condensate.$^{\ref{mr97}}$ Dynamical chiral symmetry breaking is an
essentially nonperturbative effect.

It is natural to explore DCSB using QCD's gap equation and the evolution of
the dressed-quark mass-function in Eq.\ (\ref{Mchiral}) to a large and finite
constituent-quark-like mass in the infrared, $M(0) \sim 0.5\,$GeV, is a
longstanding prediction of such Dyson-Schwinger equation (DSE)
studies$^{\,\ref{cdragw}}$ that has recently been confirmed in simulations of
quenched lattice-QCD.$^{\ref{latticequark}}$ The features and flaws of the
approach are well known.  For example, the DSEs are a keystone in proving
renormalisability and provide a generating tool for perturbation theory.
However, they are also a collection of coupled integral equations from which
a tractable problem is only obtained by truncation.  A weak coupling
expansion, which reproduces perturbation theory, is one systematic truncation
procedure.  However, that cannot be used to study nonperturbative phenomena.

\section{Truncating the DSEs}
Fortunately a systematic alternative is available.$^{\ref{truncscheme}}$ Its
leading order term furnishes a reliable description of vector and flavour
nonsinglet pseudoscalar mesons and their
interactions,$^{\ref{truncscheme2},\ref{marisRev}}$ and the reasons for
success in these channels can be understood {\it a priori}.  The scheme is
worth illustrating.

The renormalised gap equation
\begin{equation}
S(p)^{-1} = Z_2 \,(i\gamma\cdot p + m^{\rm bm})+\, Z_1 \int^\Lambda_q \, g^2
D_{\mu\nu}(p-q) \frac{\lambda^a}{2}\gamma_\mu S(q) \Gamma^a_\nu(q,p) \,,
\label{gendse}
\end{equation}
involves: $D_{\mu\nu}(k)$, the dressed-gluon propagator; $\Gamma^a_\nu(q;p)$,
the dressed-quark-gluon vertex; $m^{\rm bm}$, the $\Lambda$-dependent
current-quark bare mass that appears in the Lagrangian; $\int^\Lambda_q :=
\int^\Lambda d^4 q/(2\pi)^4$, a translationally-invariant regularisation of
the integral, with $\Lambda$ the regularisation mass-scale; and the
quark-gluon-vertex and quark wave function renormalisation constants,
$Z_1(\zeta^2,\Lambda^2)$ and $Z_2(\zeta^2,\Lambda^2)$ respectively, which
depend on the renormalisation point, the regularisation mass-scale and the
gauge parameter.

The kernel of Eq.\ (\ref{gendse}) is formed from a product of the
dressed-gluon propagator and dres\-sed-quark-gluon vertex.  It is to this
product that a systematic truncation must be applied.  However, one must also
consider more than just the gap equation's kernel.  Chiral symmetry is
expressed via the axial-vector Ward-Takahashi identity:
\begin{eqnarray} 
P_\mu \, \Gamma_{5\mu}(k;P) & = & S^{-1}(k_+)\, i\gamma_5 + i\gamma_5 
\,S^{-1}(k_-)\,, \label{avwti} 
\end{eqnarray} 
$k_\pm = k\pm P/2$, wherein $\Gamma_{5\mu}(k;P)$ is the dressed axial-vector 
vertex.  This three-point function satisfies an inhomogeneous Bethe-Salpeter 
equation: 
\begin{equation} 
\left[\Gamma_{5\mu}(k;P)\right]_{tu}= 
Z_2 \left[\gamma_5\gamma_\mu\right]_{tu} + \int^\Lambda_q  [S(q_+) 
\Gamma_{5\mu}(q;P) S(q_-)]_{sr} K_{tu}^{rs}(q,k;P)\,, 
\end{equation}
in which $K(q,k;P)$ is the renormalised fully-amputated quark-antiquark
scattering kernel, and the colour-, Dirac- and flavour-matrix structure in
the equation is denoted by the indices $r,s,t,u$.  The Ward-Takahashi
identity, Eq.~(\ref{avwti}), means that the kernel in the gap equation and
that in the Bethe-Salpeter equation are intimately related.  Therefore a
qualitatively reliable understanding of chiral symmetry and its dynamical
breaking can only be obtained using a truncation scheme that preserves this
relation, and hence guarantees Eq.~(\ref{avwti}) without a
\textit{fine-tuning} of model-dependent parameters.

The truncation scheme introduced in Ref.\ [\ref{truncscheme}] is a
dressed-loop expansion of the dressed-quark-gluon vertices that appear in the
half-amputated dressed-quark-anti\-quark scattering matrix: $S^2 K$, a
re\-nor\-ma\-li\-sa\-tion-group in\-va\-ri\-ant.$^{\ref{truncscheme2}}$ All
$n$-point functions involved thereafter in connecting two particular
quark-gluon vertices are \textit{fully dressed}.  The effect of this
truncation in the gap equation, Eq.~(\ref{gendse}), is realised through the
following representation of the quark-gluon vertex, $ \Gamma_\mu^a =
\frac{1}{2}\lambda^a\,\Gamma_\mu$:
\begin{eqnarray} 
\nonumber 
\lefteqn{Z_1 \Gamma_\mu(k,p)   =   \gamma_\mu +  \frac{1}{2 N_c} 
\int_\ell^\Lambda\! g^2 D_{\rho\sigma}(p-\ell) 
\gamma_\rho S(\ell+k-p) \gamma_\mu S(\ell) 
\gamma_\sigma}\\ 
\nonumber &+ & \frac{N_c}{2}\int_\ell^\Lambda\! g^2\, 
D_{\sigma^\prime \sigma}(\ell) \, D_{\tau^\prime\tau}(\ell+k-p)\, 
\gamma_{\tau^\prime} \, S(p-\ell)\, 
\gamma_{\sigma^\prime}\, 
\Gamma^{3g}_{\sigma\tau\mu}(\ell,-k,k-p) \\
& + & [\ldots]\,. 
\label{vtxexpand} 
\end{eqnarray} 
Here $\Gamma^{3g}$ is the dressed-three-gluon vertex and it is apparent that
the lowest order contribution to each term written explicitly is O$(g^2)$.
The ellipsis represents terms whose leading contribution is O$(g^4)$; viz.,
the crossed-box and two-rung dressed-gluon ladder diagrams, and also terms of
higher leading-order.
 
This expansion of $S^2 K$, with its implications for other $n$-point
functions, yields an ordered truncation of the DSEs that guarantees,
term-by-term, the preservation of vector and axial-vector Ward-Takahashi
identities, a feature that has been exploited$^{\,\ref{mrt98},\ref{mishaSVY}}$
to prove Goldstone's theorem and other exact results in QCD.  It is readily
seen that inserting Eq.\ (\ref{vtxexpand}) into Eq.\ (\ref{gendse}) provides
the rule by which the rainbow-ladder truncation can be systematically
improved.

\begin{figure}[t]
\centerline{\resizebox{0.85\textwidth}{!}{\includegraphics{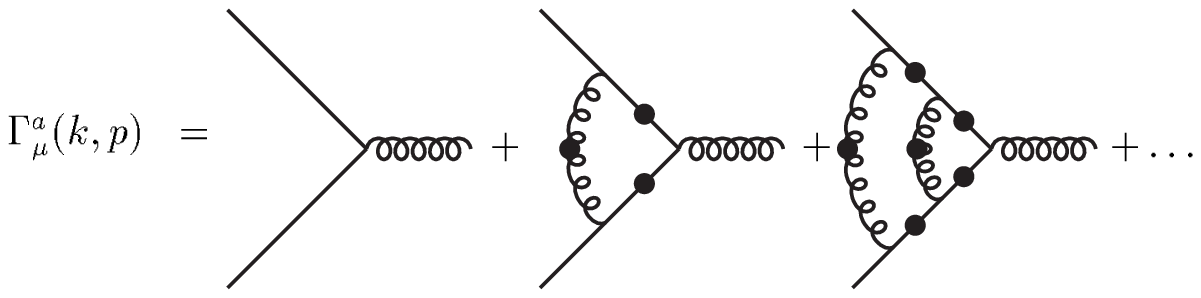}}}
\caption{\label{fig1} Integral equation for a planar dressed-quark-gluon 
vertex obtained by neglecting contributions associated with explicit gluon 
self-interactions.  Solid circles indicate fully dressed propagators.  The 
vertices are not dressed.  (Adapted from Ref.\ [\protect\ref{truncscheme2}].)} 
\end{figure}

With this scheme, as with perturbation theory, it is impossible, in general,
to obtain complete closed-form expressions for the kernels of the gap and
Bethe-Salpeter equations.  However, for the planar dressed-quark-gluon vertex
depicted in Fig.\ \ref{fig1}, closed forms can be obtained and a number of
significant features illustrated$^{\,\ref{truncscheme2}}$ when one uses the
following model for the dressed-gluon line$^{\,\ref{mn83}}$
\begin{equation} 
\label{mnmodel} g^2 \, D_{\mu\nu}(k) = 
\left[\delta_{\mu\nu} - \frac{k_\mu k_\nu}{k^2}\right] (2\pi)^4\, {G}^2
\, \delta^4(k)\,,
\end{equation} 
where $G$ sets the model's mass-scale.  This model has many positive features
and, furthermore, its particular momentum-dependence works to advantage in
reducing integral equations to character-preserving algebraic equations.
Naturally, there is a drawback: the simple momentum dependence also leads to
some model-dependent artefacts, but they are easily identified and hence not
cause for concern.
 
It is a general result$^{\,\ref{truncscheme2}}$ that with any vertex whose
diagrammatic content is known explicitly; e.g., Fig. \ref{fig1}, it is
possible to construct a unique Bethe-Salpeter kernel which ensures the
Ward-Takahashi identities are fulfilled: that kernel is necessarily
nonplanar.  This becomes transparent with the model in Eq.\ (\ref{mnmodel}),
using which the gap equation obtained with the vertex depicted in Fig.\
\ref{fig1} reduces to an algebraic equation, irrespective of the number of
dressed-gluon rungs that are retained, and the same is true of the
Bethe-Salpeter equations in every channel: pseudoscalar, vector, etc.
 
\begin{table}[b] 
\tbl{Calculated $\pi$ and $\rho$ masses, in GeV, quoted with ${G}= 0.48\,{\rm
GeV}$.  $n$ is the number of dressed-gluon rungs retained in the planar
vertex, see Fig.~\protect\ref{fig1}, and hence the order of the
vertex-consistent Bethe-Salpeter kernel: the rapid convergence of the kernel
is evident from the tabulated results.  (Adapted from Ref.\
[\protect\ref{truncscheme2}].)}  {\footnotesize
\begin{tabular*} 
{\hsize}{|l@{\extracolsep{0ptplus1fil}} 
|c@{\extracolsep{0ptplus1fil}}c@{\extracolsep{0ptplus1fil}} 
c@{\extracolsep{0ptplus1fil}}c|@{\extracolsep{0ptplus1fil}}} 
\hline
%
 & $M_H^{n=0}$ & $M_H^{n=1}$ & $M_H^{n=2}$ & $M_H^{n=\infty}$\\\hline 
$\pi$, $m=0$ & 0 & 0 & 0 & 0\\ 
$\pi$, $m=0.011$ & 0.152 & 0.152 & 0.152 & 0.152\\\hline 
$\rho$, $m=0$ & 0.678 & 0.745 & 0.754 & 0.754\\ 
$\rho$, $m=0.011$ & 0.695 & 0.762 & 0.770 & 0.770 \\ 
%
\hline 
\end{tabular*} \label{tab1} }
\end{table} 
 
Results for the $\pi$ and $\rho$ are illustrated in Table \ref{tab1}.  It is
apparent that, irrespective of the order of the truncation; i.e., the value
of $n$, the number of dressed gluon rungs in the quark-gluon vertex, the pion
is massless in the chiral limit.  (NB.\ This pion is composed of heavy
dressed-quarks, as is evident in the calculated scale of the dynamically
generated dressed-quark mass function: $M(0) \approx 0.5\,$GeV.) The
masslessness of the $\pi$ is a model-independent consequence of consistency
between the Bethe-Salpeter kernel and the kernel in the gap equation.
Furthermore, the bulk of the $\rho$-$\pi$ mass splitting is present in the
chiral limit and with the simplest ($n=0$; i.e., rainbow-ladder) kernel,
which shows that this mass difference is driven by the DCSB mechanism: it is
not the result of a finely adjusted hyperfine interaction.  Finally, the
quantitative effect of improving on the rainbow-ladder truncation; i.e.,
including more dressed-gluon rungs in the gap equation's kernel and
consistently improving the kernel in the Bethe-Salpeter equation, is a 10\%
correction to the vector meson mass.  Simply including the first correction
($n=1$; i.e., retaining the first two diagrams in Fig.\ \ref{fig1}) gives a
vector meson mass that differs from the fully resummed result by
$\stackrel{<}{\mbox{\tiny$\sim$}}1$\%.

While these results were obtained with a rudimentary interaction model, the
procedure is completely general.  However, the algebraic simplicity of the
analysis is naturally peculiar to the model.  With a more realistic
interaction, the gap and vertex equations yield a system of twelve coupled
integral equations.  The Bethe-Salpeter kernel for any given channel then
follows as the solution of a determined integral equation.

By identifying the rainbow-ladder truncation as the lowest order in a
systematic scheme the procedure also provides a means of anticipating the
channels in which that truncation must fail.  The scalar mesons are an
example.  Parametrisations of the rainbow-ladder truncation, fitted to $\pi$,
$\rho$ observables, yield scalar mesons masses that are too
large.$^{\ref{jain}}$ That was thought to be a problem.  However, we now know
this had to happen because cancellations that occur between higher order
terms in the pseudoscalar and vector channels, thereby reducing the magnitude
of corrections, do not occur in the scalar channel, wherein the full kernel
contains additional attraction.$^{\ref{pmpipi}}$ Therefore an interaction
model employed in rainbow-ladder truncation which simultaneously provides a
good description of scalar and pseudoscalar mesons must contain spurious
degrees of freedom.  Quantitative studies of the effect of the higher-order
terms have begun$^{\,\ref{pmpipi},\ref{xbox}}$ and a straightforward
understanding of scalar mesons is not impossible.  Similarly, it may be that
the only thing remarkable about mesons with ``exotic'' quantum numbers is
that their accurate description requires only an equally carefully considered
application of this systematic approach.$^{\ref{exotics}}$

\section{Concerning the quark condensate} 
While Eq.\ (\ref{Mchiral}) is expressed in Landau gauge, the so-called OPE
condensate, $\langle \bar q q \rangle^0$, is gauge parameter independent.  In
the chiral limit this condensate sets the scale of the mass function in the
ultraviolet and thus plays a role analogous to that of the
renormalisation-group-invariant current-quark mass in the massive theory.  A
determination of the OPE condensate directly from lattice-QCD simulations
must await an accurate chiral extrapolation$^{\,\ref{latticequarkchiral}}$
but DSE models tuned to reproduce modern lattice data give$^{\,\ref{raya}}$
$|\langle \bar q q \rangle^0| \sim \Lambda_{\rm QCD}^3$.

Another view of DCSB is obtained by considering the eigenvalues and
eigenfunctions of the anti-Hermitian massless Euclidean Dirac operator:
\begin{equation} 
\gamma\cdot D \, u_n(x) = i \lambda_n \, u_n(x)\,.
\end{equation} 
The eigenfunctions form a complete set, and except for zero modes they occur
in pairs: $\{u_n(x), \gamma_5 u_n(x)\}$, with eigenvalues of opposite sign.
It follows that in an external gauge field, $A$, one can write the quark
Green function
\begin{equation} 
S(x,y;A) = \langle q(x) \bar q(y) \rangle_A^m = \sum_n \frac{u_n(x) \, 
u_n^\dagger(y)}{i \lambda_n + m}\,, 
\end{equation} 
where $m$ is the current-quark mass.  Assuming, e.g., a lattice
regularisation,
\begin{equation} 
\label{qbqonV} 
\frac{1}{V} \, \int_V d^4x \, \langle \bar q(x) q(x) \rangle_A^m = - \frac{2
m}{V} \, \sum_{\lambda_n>0} \, \frac{1}{\lambda_n^2 + m^2}\,,
\end{equation} 
where $V$ is the lattice volume.\footnote{In deriving Eq.\
(\protect\ref{qbqonV}), zero modes have been neglected, which is justified
under broad conditions.$^{\protect\ref{leutwyler}}$} One may now define a
quark condensate:
\begin{equation} 
\label{BCqbq1} 
\langle 0 | \bar q q | 0 \rangle^m := \lim_{V\to \infty} \frac{1}{V} \,
\int_V d^4x \, \left\langle \, \langle \bar q(x) q(x) \rangle_A^m
\,\right\rangle\,,
\end{equation} 
wherein the r.h.s.\ expresses an average over gauge field configurations.
When $V\to \infty$ the operator spectrum becomes dense and Eqs.\
(\ref{qbqonV}), (\ref{BCqbq1}) become
\begin{equation} 
\label{bcone} 
- \langle 0 | \bar q q | 0 \rangle^m = 2 m \int_0^\infty d\lambda\, 
\frac{\rho(\lambda)}{ \lambda^2 + m^2}\,, 
\end{equation} 
with $\rho(\lambda)$ a spectral density.  This equation expresses an
assumption that in QCD the full two-point massive-quark Schwinger function,
when viewed as a function of the current-quark mass, has a spectral
representation.
 
It follows formally from Eq.\ (\ref{bcone}) that 
\begin{equation} 
\label{bcrelation}
- \langle 0 | \bar q q | 0 \rangle^0:= \lim_{m\to 0} \, 2 m \int_0^\infty
d\lambda\, \frac{\rho(\lambda)}{ \lambda^2 + m^2} = \pi \, \rho(0)\,.
\end{equation}  
This is the so-called Banks-Casher relation,$^{\ref{bankscasher}}$ which has
long been advocated as a means by which a quark condensate may be measured in
lattice-QCD simulations.$^{\ref{Marinari}}$

The explication$^{\ref{langfeld}}$ of a correspondence between the condensate
in Eq.\ (\ref{Mchiral}) and that in Eq.\ (\ref{bcrelation}) requires care
because Eq.\ (\ref{bcone}) is meaningless as written: dimensional counting
reveals the r.h.s.\ has mass-dimension three and since $\lambda$ will at some
point be greater than any relevant internal scale, the integral must diverge
as $\Lambda^2$, where $\Lambda$ is the regularising mass-scale.

The Schwinger function
\begin{equation} 
\label{defsigmam} 
\tilde \sigma(m) := N_c\,{\rm tr}_D\int_p^\Lambda\! \tilde S_m(p) \,, 
\end{equation}
which is the trace of the unrenormalised massive dressed-quark propagator
evaluated at a fixed value of the regularisation scale, $\Lambda$, can be
identified with the l.h.s.\ of Eq.\ (\ref{bcone}).  The essence of the
Banks-Casher relation is that $\tilde\sigma(m)$ have a spectral
representation:
\begin{equation} 
\label{defsigmamspectral} 
\tilde \sigma(m) := 2 m \, \int_0^\Lambda \! d \lambda \frac{\tilde 
\rho(\lambda)}{\lambda^2 + m^2}\,, 
\end{equation} 
where $m = m^{\rm bm}(\Lambda)$, and Eq.\ (\ref{defsigmamspectral}) entails
\begin{equation} 
\label{discty} 
\tilde \rho(\lambda) = \frac{1}{2 \pi} \lim_{\eta\to 0^+} \left[ \, \tilde 
\sigma(i \lambda + \eta) - \tilde \sigma(i \lambda - \eta)\, \right]\,. 
\end{equation}  

The content of this sequence of equations is elucidated by inserting the free
quark propagator in Eq.\ (\ref{defsigmam}).  The integral thus obtained is
readily evaluated using dimensional regularisation:
\begin{equation} 
\tilde \sigma_{\rm free}(m) = \frac{N_c}{4\pi^2} \, m^3 \left[ 
\ln\frac{m^2}{\zeta^2}+ \frac{1}{\varepsilon} + \gamma - \ln 4\pi \right]. 
\end{equation} 
With Eq.\ (\ref{discty}) the regularisation dependent terms cancel and one 
obtains 
\begin{equation} 
\label{rhopert} 
\tilde \rho(\lambda) = \frac{N_c}{4\pi^2} \, \lambda^3\,. 
\end{equation} 
In perturbation theory it can be shown that every contribution to $\tilde
\rho(\lambda)$ is proportional to $\lambda^3$ and hence
\begin{equation} 
\langle 0 | \bar q q | 0 \rangle^0 \propto \tilde \rho(\lambda=0) = 0 \,.
\end{equation} 
A nonzero value of $\rho(0)$ is an essentially nonperturbative effect.
 
\begin{figure}[t] 
\centerline{\resizebox{0.80\textwidth}{!}{\includegraphics{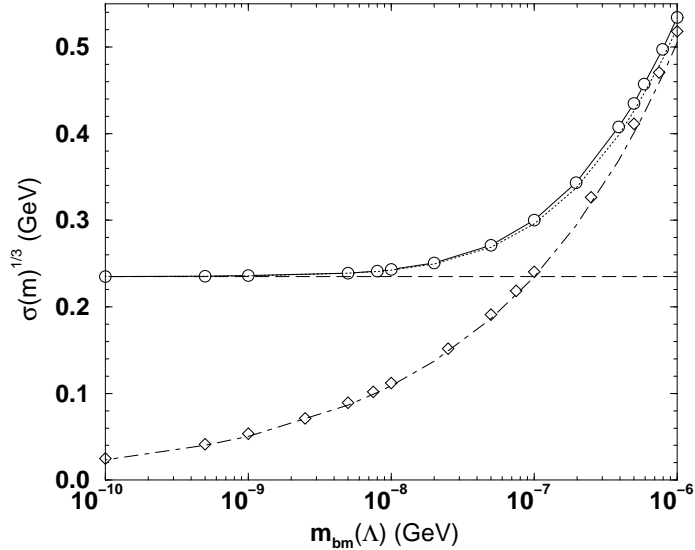}}} 
\caption{\label{fig2} Circles/solid-line: $\sigma(m)^{1/3}$ in Eq.\
(\protect\ref{sigmam}) as a function of the current-quark bare-mass,
evaluated using the dressed-quark propagator obtained in the model of Ref.\
[\protect\ref{mr97}]; dashed line: the model's value of $(-\langle \bar q q
\rangle^0_{\zeta=1\,{\rm GeV}}) = (0.24\,{\rm GeV})^3$; and dotted line: Eq.\
(\protect\ref{sigmamexpected}).  Diamonds: $\sigma(m)^{1/3}$ evaluated in a
non-confining version of the model; dot-dashed line: Eq.\
(\protect\ref{sigmamexpected}) with $\langle \bar q q \rangle^0 \equiv 0$.
(Adapted from Ref.\ [\protect\ref{langfeld}].)}
\end{figure}

A precise analysis requires that attention be paid to renormalisation.
Therefore consider the gauge-parameter-independent trace of the renormalised
massive quark propagator:
\begin{eqnarray} 
\nonumber \sigma(m;\zeta) & = & Z_4(\zeta,\Lambda)\, N_c\,{\rm
tr}_D\int_p^\Lambda\! S_m(p;\zeta) \,,
\label{sigmam} 
\end{eqnarray} 
where the argument remains $m=m^{\rm bm}(\Lambda)$.  The renormalisation
constant $Z_4$ vanishes logarithmically with increasing $\Lambda$ and hence
one still has $\sigma(m) \sim \Lambda^2\, m^{\rm bm}(\Lambda)$.  However,
since$^{\,\ref{mrt98}}$
\begin{equation}
- \langle \bar q q \rangle_{\zeta}^0= \lim_{\Lambda \to \infty}
 Z_4(\zeta,\Lambda)\, N_c\, {\rm tr}_D \!  \int^\Lambda_p \!  S_{0}(p;\zeta)\,,
\label{qbqzeta} 
\end{equation} 
it is clear that for any finite but large value of $\Lambda$ and tolerance
$\delta$, it is always possible to find $m_\delta^{\rm bm}(\Lambda)$ such
that
\begin{equation} 
\label{sigmabound} 
\sigma(m;\zeta) + \langle \bar q q \rangle^0_\zeta < \delta\,,\; \forall
m^{\rm bm}< m_\delta^{\rm bm}.
\end{equation} 
This is true in QCD and can be illustrated using the model of Ref.\
[\ref{mr97}].  Figure \ref{fig2} displays $\sigma(m,\zeta)$, evaluated using
a hard cutoff, $\Lambda$, on the integral in Eq.\ (\ref{sigmam}), calculated
with the massive dressed-quark propagators obtained by solving the gap
equation.  Since Eq.\ (\ref{sigmabound}) specifies the domain on which the
value of $\sigma(m;\zeta)$ is determined by nonperturbative effects, one
anticipates
\begin{equation} 
\label{epsilondomain} 
m_\delta^{\rm bm}(\Lambda) \approx -\langle \bar q q \rangle^0 / 
\Lambda^2 \sim 10^{-9} 
\end{equation} 
for $\Lambda = 2.0\,$TeV in QCD where $|\langle \bar q q \rangle^0| \sim
\Lambda_{\rm QCD}^3$.  This estimate is confirmed in Fig.\ \ref{fig2}.

The dotted line in Fig.\ \ref{fig2} is
\begin{eqnarray} 
\nonumber \lefteqn{\!\! \sigma(m,\zeta) = 
}\\ 
& & \!\!  -\langle \bar q q \rangle^0_\zeta \, 
\frac{2}{\pi} \arctan \frac{\Lambda}{m} + Z_4(\zeta,\Lambda)\, \frac{N_c}{4\pi^2} \, m\left[ \Lambda^2 - m^2\ln 
\left(1+\Lambda^2 / m^2\right)\right] ,
\label{sigmamexpected} 
\end{eqnarray} 
and the difference between this and the curve is of order $ \alpha(\Lambda)
\, m(\Lambda)\, \Lambda^2 $ because the DSE model incorporates QCD's one-loop
behaviour.  The figure also displays $\sigma(m,\zeta)$ obtained in the
absence of confinement, in which case$^{\,\ref{hawes}}$ $\langle \bar q q
\rangle^0\equiv 0$, as is apparent.
 
It is now plain that in QCD $\sigma(m,\zeta)$ is a monotonically increasing
convex-up function with a regular chiral limit and consequently it has a
spectral representation:
\begin{equation} 
\sigma(m,\zeta) = 2 m \int_0^\Lambda \frac{\rho(\lambda)}{\lambda^2 +
m^2}\,.
\end{equation} 
This lays the vital plank in a veracious connection between the condensates in 
Eqs.\ (\ref{Mchiral}) and (\ref{bcrelation}).  On the domain specified by Eq.\ 
(\ref{epsilondomain}), the behaviour of $\sigma(m,\zeta)$ in Eq.\ 
(\ref{sigmam}) is given by Eq.\ (\ref{sigmamexpected}), which yields, via Eq.\ 
(\ref{discty}), 
\begin{equation} 
\label{finalrho} 
\pi\, \rho(\lambda) = -\langle \bar q q \rangle^0_\zeta +
Z_4(\zeta,\Lambda)\, \frac{N_c}{4\pi} \, \lambda^3 + \ldots \,,
\end{equation} 
where the ellipsis denotes contributions from the higher-order terms implicit 
in Eq.\ (\ref{sigmamexpected}). 

Reference [\ref{Harald}] reports the spectral density of the staggered Dirac
operator in quenched $SU(3)$ gauge theory calculated on a $V = 4^4$-lattice,
in the vicinity of the deconfining phase transition at $\beta \gtrsim 5.6$.
A comparison$^{\,\ref{langfeld}}$ with these results shows that while the
effect of finite lattice volume is evident early (for $\lambda a \gtrsim
0.1$, $a$ is the lattice spacing) the behaviour at small $\lambda a$ is
qualitatively in agreement with Eq.\ (\ref{finalrho}) and Fig.\ \ref{fig2}: a
nonzero OPE condensate dominates the Dirac spectrum in the confined domain;
and it vanishes in the deconfined domain, whereupon $\rho(0)=0$ and the
perturbative evolution, Eq.\ (\ref{rhopert}), is manifest.  To be more
quantitative, for a lattice coupling $\beta = 5.4$, $\rho(0)a^3 \approx 70$,
so that
\begin{equation} 
\label{pirho0lattice} \pi \rho(0)\, a^3/V \approx (0.95)^3. 
\end{equation} 
The lattice spacing was not determined in Ref.\ [\ref{Harald}] but one can
nevertheless assess the scale of Eq.\ (\ref{pirho0lattice}) by supposing $a
\sim 0.3\, {\rm fm} \sim 0.3/ \Lambda_{\rm QCD}$, a value not unreasonable
for small lattices and small $\beta$, wherewith the r.h.s.\ is $\sim (3
\Lambda_{\rm QCD})^3\!$. This is too large but not unreasonable given the
parameters of the simulation, its errors and the systematic uncertainties in
our estimate.  Fitting the lattice data at $\beta = 5.8$, one finds
$\rho(\lambda) \propto \lambda^3 $ on $\lambda< 0.1$ but with a
proportionality constant larger than that anticipated from perturbation
theory; viz.\ Eq.\ (\ref{finalrho}).  Some mismatch was to be expected,
however, because at $\beta=5.8$ one has only just entered the deconfined
domain and near the transition boundary some nonperturbative effects are
still relevant. It is a modern challenge to determine those gauge couplings
and lattice parameters for which the data are quantitatively consistent with
Eq.\ (\ref{finalrho}).

In practice, there are three main parameters in a simulation of lattice-QCD:
the lattice volume, characterised by a length $L$; the lattice spacing, $a$;
and the current-quark mass, $m$.  So long as the lattice size is large
compared with the current-quark's Compton wavelength; viz., $L\gg 1/m$, then
dynamical chiral symmetry breaking can be expressed in the
simulation. Supposing that to be the case then, as outlined above, so long as
the lattice spacing is small compared with the current-quark's Compton
wavelength
\begin{equation} 
\pi \, \rho(0)  \approx-\langle \bar q q \rangle^0_{1/a} \,,\; a\ll 1/m\ll L \,,
\end{equation} 
where the $\langle \bar q q \rangle^0_{1/a}$ is the scale-dependent OPE
condensate [$\zeta = \Lambda = 1/a$ in Eq.\ (\ref{finalrho})].  This
completes an explication of the connection between the condensates.  Note,
however, that the continuum analysis indicates that one requires $a m
\lesssim (a \Lambda_{\rm QCD})^3$ if $\rho(\lambda= 0)$ is to provide a
veracious estimate of the OPE condensate.  The residue at the lowest-mass
pole in the flavour-nonsinglet pseudoscalar vacuum polarisation provides a
measure of the OPE condensate that is accurate for larger current-quark
masses.$^{\ref{langfeld}}$

\section*{Acknowledgments} 
I am grateful to the Organising Committee for the opportunity to participate
in the ``5th International Conference on Quark Confinement and the Hadron
Spectrum.''  This work was supported by: the Department of Energy, Nuclear
Physics Division, contract no.~\mbox{W}-31-109-ENG-38; the National Science
Foundation, grant no.\ INT-0129236; and benefited from the resources of the
National Energy Research Scientific Computing Center.

\section*{References} 
\begin{small}
\begin{enumerate} 
\setlength{\baselineskip}{2.6ex}
\setlength{\parskip}{0.1ex}
\item\hspace{-0.5em}.\ P.\ Maris and C.D.\ Roberts, {\it Phys.\
Rev.} {\bf C56}, 3369 (1997). \label{mr97}
 
\item\hspace{-0.5em}.\  C.D.~Roberts and A.G.~Williams, 
{\it Prog.\ Part.\ Nucl.\ Phys.}  {\bf 33}, 477 (1994). \label{cdragw}

\item\hspace{-0.5em}.\ P.O.\ Bowman, U.M.\ Heller and A.G.\ Williams,
{\it Phys.\ Rev.} {\bf D66}, 014505 (2002). \label{latticequark}

\item\hspace{-0.5em}.\  A.\ Bender, C.D.\ Roberts and L.\ v.\ Smekal, 
{\it Phys.\ Lett.} {\bf B380}, 7 (1996). \label{truncscheme}

\item\hspace{-0.5em}.\  A.\ Bender, W.\ Detmold, A.W.\ Thomas and C.D.\ Roberts,
{\it Phys.\ Rev.} {\bf C65}, 065203 (2002). \label{truncscheme2}

\item\hspace{-0.5em}.\  P.\ Maris and C.D.\ Roberts, ``Dyson-Schwinger Equations:
A tool for Hadron Physics,'' nucl-th/0301049. \label{marisRev}

\item\hspace{-0.5em}.\ P.\ Maris, C.D.\ Roberts and P.C.\ Tandy, {\it Phys.\
Lett.} {\bf B420}, 267 (1998). \label{mrt98}

\item\hspace{-0.5em}.\ M.A.\ Ivanov, Yu.L.\ Kalinovsky and C.D.\ Roberts,
{\it Phys.\ Rev.} {\bf D60}, 034018 (1999). \label{mishaSVY}

\item\hspace{-0.5em}.\ H.J.\ Munczek and A.M.\ Nemirovsky, 
{\it Phys.\ Rev.} {\bf D28}, 181 (1983). \label{mn83}

\item\hspace{-0.5em}.\ P.\ Jain and H.J.\ Munczek, {\it Phys.\ Rev.}  {\bf
D48}, 5403 (1993); \label{jain}
P.\ Maris, C.D.\ Roberts, S.M.\ Schmidt and P.C.\ Tandy,
{\it Phys.\ Rev.} {\bf C63}, 025202 (2001);
P.\ Maris, 
{\it Few Body Syst.}  {\bf 32}, 41 (2002). 

\item\hspace{-0.5em}.\ S.R.\ Cotanch and P.\ Maris, ``QCD based quark
description of $\pi - \pi$ scattering up to the $\sigma$ and $\rho$ region,''
hep-ph/0210151. \label{pmpipi}

\item\hspace{-0.5em}.\ A.\ H\"oll, P.\ Maris and C.D.\ Roberts, 
{\it Phys.\ Rev.} {\bf C59}, 1751 (1999);
J.C.R.\ Bloch, C.D.\ Roberts and S.M.\ Schmidt, {\it Phys.\ Rev.}  {\bf C60},
065208 (1999);
P.\ Maris and P.C.\ Tandy, {\it Phys.\ Rev.} {\bf C62}, 055204 (2000);
A.\ Bender, W.\ Detmold and A.W.\ Thomas, 
{\it Phys.\ Lett.} {\bf B516}, 54 (2001). \label{xbox}

\item\hspace{-0.5em}.\ J.C.R.\ Bloch, Yu.L.\ Kalinovsky, C.D.\ Roberts and
S.M.\ Schmidt,
{\it Phys.\ Rev.} {\bf D60}, 111502 (1999);
C.J.\ Burden and M.A.\ Pichowsky,
{\it Few Body Syst.}  {\bf 32}, 119 (2002).
\label{exotics}

\item\hspace{-0.5em}.\ J.B.\ Zhang, F.D.R.\ Bonnet, P.O.\ Bowman, D.B.\
Leinweber and A.G.\ Williams, ``Towards the Continuum Limit of the Overlap
Quark Propagator in Landau Gauge,'' hep-lat/0208037. \label{latticequarkchiral}

\item\hspace{-0.5em}.\ P.\ Maris, A.\ Raya, C.D.\ Roberts and S.M.\ Schmidt,
``Confinement and dynamical chiral symmetry breaking,''
nucl-th/0208071. \label{raya}

\item\hspace{-0.5em}.\ H.\ Leutwyler and A.\ Smilga,
{\it Phys.\ Rev.} {\bf D46}, 5607 (1992). \label{leutwyler}
 
\item\hspace{-0.5em}.\ T.\ Banks and A.\ Casher,
{\it Nucl.\ Phys.} {\bf B169}, 103 (1980).\label{bankscasher}

\item\hspace{-0.5em}.\ E.\ Marinari, G.\ Parisi and C.\ Rebbi,
{\it Phys.\ Rev.\ Lett.}  {\bf 47}, 1795 (1981). \label{Marinari}

\item\hspace{-0.5em}.\ K.\ Langfeld, R.\ Pullirsch, H.\ Markum, C.D.\ Roberts
and S.M.\ Schmidt, ``Concerning the quark condensate,''
nucl-th/0301024. \label{langfeld}

\item\hspace{-0.5em}.\ A.\ Bender, D.\ Blaschke, Yu.L.\ Kalinovsky and C.D.\
Roberts,
{\it Phys.\ Rev.\ Lett.}  {\bf 77}, 3724 (1996);
F.T.\ Hawes, P.\ Maris and C.D.\ Roberts, 
{\it Phys.\ Lett.} {\bf B440}, 353 (1998). \label{hawes}

\item\hspace{-0.5em}.\ E.\ Bittner, H.\ Markum and R.\ Pullirsch,
{\it Nucl.\ Phys.\ Proc.\ Suppl.}  {\bf 96}, 189 (2001). \label{Harald}
 
\end{enumerate}
\end{small}

\end{document}